# Examining the social aspects of Enterprise Architecture Implementation: A Morphogenetic Approach

**Edi Nuryatno**
School of Business
Edith Cowan University
Perth, Western Australia
Email: e.nuryatno@ecu.edu.au

**Philip Dobson**
School of Business
Edith Cowan University
Perth, Western Australia
Email: p.dobson@ecu.edu.au

## Abstract

This paper argues that existing research in Enterprise Architecture (EA) tends to emphasise the technical aspects of implementation and neglects the role of 'people'. The paper uses Archer's Morphogenetic Approach (MA) to elaborate the social context of EA. This approach examines the relationships between agency, structure and culture to understand how stakeholders affect and are affected by EA implementation. A university just commencing an EA program is used as a case example. The case example describes the challenging university environment and provides an illustration of the important situational logics that direct agency action within the complex social context of a university.

## 1  Introduction

Enterprise architecture (EA) can be seen as the strategic context for the evolution of the organisational Information Systems (IS) in response to the constantly changing needs of the business environment. It thus describes organisational plans, visions, objectives and problems and the information required to support organisational goals. Recent research has found that 'people' have been ranked as a major problem in EA implementation (Raadt et al. 2008; Gartner 2009; Janssen 2012). It is suggested that the lack of focus on the 'people' aspects of EA could be the reason why many organisations still struggle with EA implementation. As Bente et al. (2012) suggests EA "…deals with social elements such as collaborative business processes, organisational leadership, political dynamics and work culture…" (p. 36) and thus requires a careful examination of the role of people: "The people element brings complex behavioural attributes into the functioning of an enterprise…" (p. 35). Such a perspective is core within Jansen's (2012) notion of the socio-political within his description of EA as: "…a means to inform, guide, direct, and constrain the decisions taken by human beings within organizations" (p. 25). In order to address this need for a greater recognition of the role of people and the social aspects of EA implementation this paper suggests the adoption of critical realism and its most recognised methodological complement the morphogenetic approach (MA) (Archer, 1995; 2013) as a useful tool to research the topic. Such a perspective suggests an analytical separation between culture, structure and agency in order to examine their interactions over time – it sees the sociocultural consequences of an interaction between the cultural and the structural to provide particular situational logics that direct but do not determine agents' actions. Under the stratified view of agency employed by Archer (1995) 'people' are separately described as human beings, actors and agents with particular institutional roles and positions. As will be demonstrated such a perspective provides a fertile grounding for examination of EA implementation.

## 2  Research context, problem and question

In the past two decades, the Higher Education sector in Australia has moved towards corporatisation, marketization and rationalisation (Gengatharen et al, 2009) as universities face significant challenges due to market change, new teaching models, and the need for efficiency savings, all of which require changes in technology capability. It is in this climate that universities encounter strategic and operational challenges that require agile IS to respond quickly to meet new expectations of students, staff and other stakeholders while facing ever-increasing costs pressure (Anderson and Backhouse 2009). EA seeks to address these challenges by providing a holistic view of the planning and development of an organisation's business, application and technology architecture.

The importance of people in EA implementation will require acceptance of the EA implementation as a social program heavily influenced by the social and cultural systems surrounding the architecture. It will also be heavily impacted by the belief structures underlying the program implementation – the "theory" or mechanisms built in to the new program. As Astbury and Leeuw (2010) suggest understanding program success and failure requires an understanding of both the underlying theory inbuilt in to the program and the context within which it is implemented. EA methodologies and frameworks have particular theories underlying their largely technical focus. It is important to understand such theories as well as the social context of their implementation. It is the argument of this paper that such frameworks need to better recognize the social aspects of their target audience.

As Astbury and Leeuw (2010) suggest it is only by understanding both the theories underlying programs and the social context within which they are implemented can we determine "…how and why programs work (or fail to work) in different contexts and for different program stakeholders." (p. 364). They fix on defining mechanisms as "…underlying entities, processes, or structures which operate in particular contexts to generate outcomes of interest" (p. 368). Mechanisms are seen as sensitive to variations in context, since in a particular context a program mechanism (or theory) may or may not be activated due to contingent conditions or possible countervailing mechanism(s). Mechanisms will play an important role in this research and can best be seen as "triggerable causal powers" (Mason et al, 2013). To explore the interplay between social structures, socio-cultural system and agency, as a first step, we seek to determine the deep causal mechanisms that constrain and enable social acceptance of EA. The main research question that this research seeks to answer are "What social mechanisms might constrain and enable the success of the EA program at university X? What are the key implementation mechanisms behind the EA program?"

## 3　Research philosophy and its practical social theories

The adoption of critical realism has significant implications with respect to the objects to be investigated, the progress of subsequent research and the outcomes that can be expected. The methodological approach used will be Archer's Morphogenetic Approach (MA). Such an approach provides the important link between a realist ontology and practical social outcomes and provides a basis for a consistency between ontology and methodology. In order to better understand the research process it is important to have an understanding of critical realism (CR).

### 3.1　Why critical realism?

Bhaskar's (1978) concept of CR distinguishes among three ontological domains in reference to social reality: the empirical, the actual and the real. The empirical domain consists of events that are actually perceived or experienced directly or indirectly, whilst the actual domain includes those events whether experienced or not. Both are encompassed by the real domain which is made up of structures and mechanisms that are relatively enduring with potential powers and properties that are activated or triggered in particular contexts or by agency action and thus may be causal in generating perceived or non-perceived events.

Bhaskar (1978, p. 25) proposes that events or phenomena should not be the core focus of research, instead the focus should be on the structures and mechanisms that generate phenomena. Applying such a focus to the examination of the EA implementation will require an understanding of the social structures and mechanisms that are currently in place and how the individual agents react to the new impositions both in terms of the increased governance role and also the impact on the way that they currently do things.

### 3.2　Critical realism practical social theories: Analytical dualism and the morphogenetic approach

In adopting CR there are a number of philosophical assumptions that must be met; one of the most important is the analytical separation of culture, structure and agency over the time. This is what Archer (1989) terms as analytical dualism, which is a fundamental component of the critical realist approach. The aim of analytical dualism is to make possible an examination of the complex duality of structure and agency – it is an artificial separation or dualism to allow the examination of a complex duality between structures (macro) and agents (micro). It suggests:

a) Structure necessarily pre-dates the action(s) leading to its reproduction or transformation, while

b) Structural elaboration necessarily post-dates the action sequences which gave rise to it (Archer, 1995).

The morphogenetic approach (morphogenesis/morphostatis: MA) is a meta-theoretical social ontology developed by Archer (1979; 1989; 1995; 2013) and serves as a methodological complement of CR by application of an analytical dualism over time. Morphogenesis 'refers to the complex interchanges that produce change in a system's given form, structure or state ("morphostasis" is the reverse)' (Archer 1989, p. xxii). The MA analysis works in terms of three part cycles: (a) structural conditioning, which refers to pre-existing structures that condition but do not determine, (b) social interaction, which arises from actions oriented towards the realisation of interests and needs emanating from current agents and may lead to, (c) structural elaboration or modification, that is, a change in the relations between the parts of the social system.

Archer views structure, culture and agency to be analytically distinct strata of social reality in which structures are viewed as "…relatively enduring, anterior social objects that possess causal powers and are neither observable nor reducible to social interaction" (Luckett, 2012, p. 340). Archer (1995) presents the social world as a stratified model involving: (a) different structural emergent properties (SEPs); roles, institutional structures, social systems, and positions, (b) different cultural emergent properties (CEPs); such as ideas, beliefs, values and ideologies, and (c) different people emergent properties (PEPs) dependent on a stratified model of human beings, actors and agents.

Archer suggests that the emergent properties of collectivities and individuals differ from the emergent properties of organised groups, which differ yet again from those pertaining to populations (p. 190). Yet as detailed in Figure 1 "these different levels of 'social integration' are not discrete from the powers of 'system integration', despite their capacity for independent variation at any given time" (p. 190). Archer describes a double morphogenesis:

"…where agency undergoes transformation, acquiring new emergent powers in the very process of seeking to reproduce and transform structures. For in such structural and cultural struggles, consciousness is raised as collectivities are transformed from primary agents into promotive interest groups; social selves are re-constituted as actors personify roles in particular ways to further their self-defined ends; and corporate agency is re-defined as institutional interests promote reorganization and re-articulation of goals in the course of strategic action for their promotion or defence" (p. 190-191).

This stratified representation of Agents is described on the right hand side of Figure 1 and allows a rich representation of the people role in organisational change as they can be seen to provide primary agency in particular positions, corporate agency in institutions or as individual actors in particular roles.

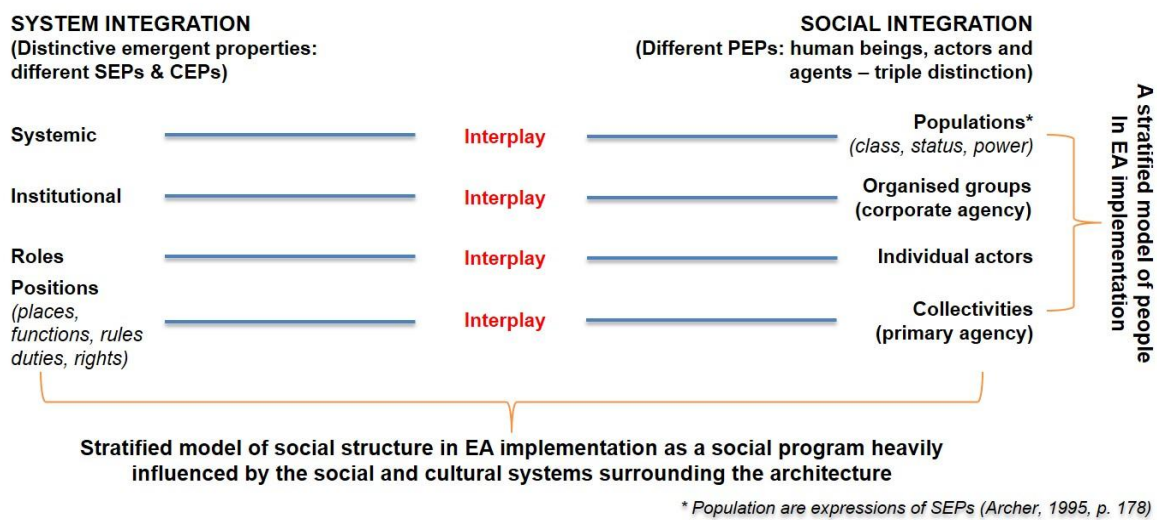

Figure 1: *Analytical dualism in social theory: A stratified model of social structure involving SEPs, CEPs and PEPs in EA implementation (adapted from Archer, 1995, p. 190)*

## 3.3　Situational logics and EA pathways

Archer (1988, 1995) argues that at the institutional level, interactions between emergent structural and cultural properties create different situational logics which predispose agents to follow particular courses of action to promote their important personal projects. Archer argues that within the cultural system there may be consistency or contradiction between ideas. Consistency and contradiction can be necessary (internally related as the ideas depend on each other and cannot operate apart) or contingent (externally related and contextual).

As described in Figure 2 Archer describes the causal influences exerted by the cultural system on the socio-cultural and defines 4 possible interactions between the cultural ("A") and the structural ("B") systems:

a) *Necessary complementarities* – The ideational compatibilities between A and B lead to an environment of mutual support. "In other words invoking A also ineluctably evokes B, but since the B upon which this A depends is consistent with it, then B buttresses adherence to A." (p. 234). The structural and the ideational are in harmony and such a position has causal possibilities at a structural level – creating a situational logic of protection at the systems level. Archer suggests that increasing depth of systemization at the structural level increasingly blocks change because of its threatening disruptive capacity. At the personal level, corporate agents see the increasing barrier to their advancement and may thus seek unpredictable avenues to break out of the constraining systemic limitations.

b) *Necessary incompatibilities* – These are the reverse of necessary complementarities; components of the socio-cultural system contain some particular belief or theory which is internally inconsistent with ideas at the cultural level. "When the constitution of the social system is marked by incompatibilities between institutions which are none the less internally and necessarily related, this has rightly been seen as containing a potential for change which is entirely lacking in the complementary configurations. Generally, when two or more institutions are necessarily and internally related to one another yet the effects of their operations are to threaten the endurance of the relationship itself, this has been referred to as a state of 'contradiction'." (p 222). Such incompatibility, or contradiction, provides a situational logic of correction as these ideas must ultimately accommodate each other. At the structural level the need for accommodation suggests the emergence of properties directed by compromise as parties struggle to remain in power. Unification is a consequence at the socio-cultural level as compromise becomes essential and emergent. This holding state is inherently unstable and suggests a period of instability as participants jockey for position with accommodation as the focus, seeking to survive amongst the incompatible cultural ideas.

c) *Contingent incompatibilities* – occur when the material world produces situations which are incompatible with the prevailing social and cultural properties: "because partisans of A and B are unconstrained by any dependence between these items, there is nothing which restrains their combativeness for they have everything to gain from inflicting maximum damage on one another's ideas in the course of competition" (Archer, 1988, p. 240). A "battleground of ideas" emerges providing a situational logic of elimination; the pluralism at a cultural level promotes the creation of distinct loyalties at a socio-cultural level – such cleavage encouraging competition at a structural level and polarization at a social level as groups and individuals struggle to remain in the game. Cooperation on the acceptance of change is discouraged and diversity reduced as elimination of alternatives is attempted and progressed.

d) *Contingent complementarities* – occur when material opportunities arise that resonate with the social and cultural properties, stimulating opportunism - "Only the contingent complementarity simultaneously holds out choices to the adherents of A but leaves them free to make what they will (if anything) of B…only the contingent compatibility is free from sociocultural manipulation, designed to induce avoidance or adoption or aversion. Certainly, distracting sociocultural practices – habitual preoccupations, established routines, traditional preserves or conventional divisions of subjects – may well reduce subjective willingness to explore new and congruent possibilities, but these will usually coexist with various sticks and carrots which stimulate originality, innovation and experimentation (as in the derived sequence). The actors concerned have substantial freedom to survey or to ignore the broader horizon which has come in to view…" (Archer 1988, p. 243). The situational logic of opportunism has a net systemic result of great cultural variety, it "breaks down artificial knowledge barriers, stimulates new departures and bold syntheses". At the cultural level wild ideas and daring proposals can ensue unchecked by the socio-cultural. At the socio-cultural diversification, specialization and recombination can ensue as "marginals disengage themselves to recoalesce in a group with a novel brief" (p. 244).

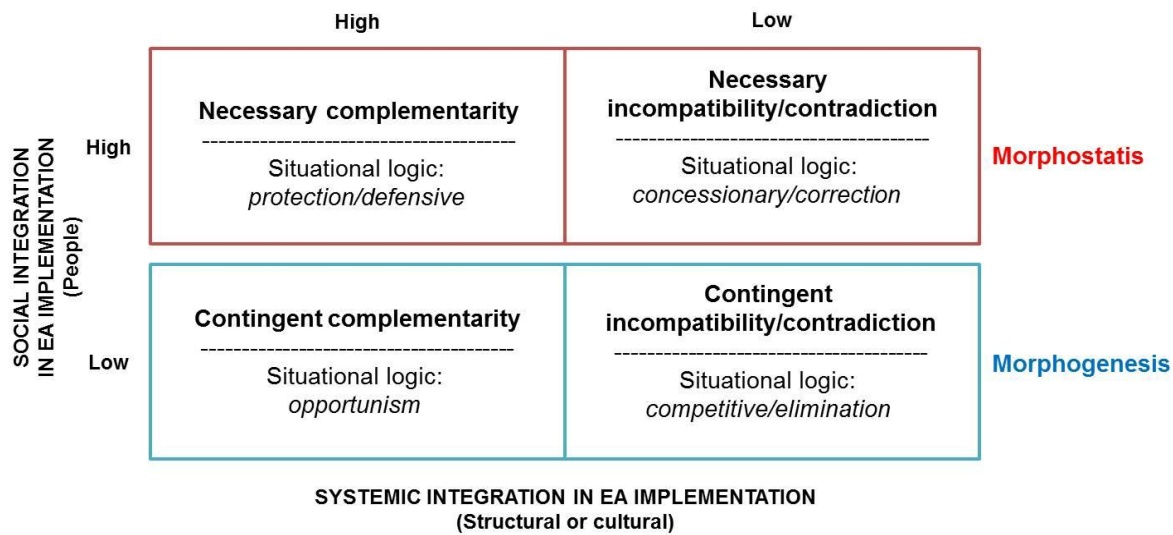

*Figure 2: Morphostatis vs morphogenesis: Situational logics in social and system integration in EA implementation (adapted from Archer, 1995, pp. 218 and 295)*

The MA approach can be applied at the level of society as proposed by Archer (2013) or it can be applied at lower levels. Applying the MA to the internal university environment (for an example see Luckett 2012) we can examine the roles of the different centres within the overall university system. At this level one could argue that the internal operation of the university today can be framed as an environment of necessary incompatibilities in that the cultural requirement of academic freedom conflicts with a systemic need for an increased managerialism and greater control; the EA implementation providing the battleground for such contradiction in that the EA reflects the high-level university requirement for greater control and centralization of the IT function.

EA, in seeking to evangelize the corporate commitment to control, impinges on the centres desire for freedom to do what they wish. As detailed in Figure 3 the situational logic of correction suggests an outcome at the cultural ideational level of syncretism ("that is, the attempt to sink differences and effect union between the contradictory elements concerned" p. 233). This situational logic must lead to compromise at a systemic level as one or both parties lead to a re-definition of the EA to accommodate the irreconcilable ideational differences. The main thrust of this situational logic is the "sinking of differences" and a unification and compromise. Archer does not include time in the detailed figure but it is implicit in the development of the argument.

The state of contradiction would hopefully be short-lived as the benefits achievable by compromise become clear to the parties imposed on. An alternative pathway involving contingent incompatibilities may ensue as parties deem unification impossible and unnecessary. Under this situational logic a battle must rage as logic moves towards elimination, accepting the divergence at a cultural level and leading to cleavage at the socio-cultural level and competition at a systemic level. The battle rages and centres seek advantage; power and politics playing a crucial and important role. Such cleavage is not a desirable option given the immediate threats evident within the university environment. Such an un-productive environment cannot be allowed to ensue as the universities do not have the luxury of time due to the urgency of imposing external issues.

Yet as Archer suggests the acceptance or otherwise of these logics depends on their ultimate social reception. Agents are define by Archer as "collectivities sharing the same life chances" – everyone is inescapably an agent in some of their doings by being part of a collective, intentionally or otherwise. Archer distinguishes between Corporate agents and Primary agents. Primary agents are those who have no say in possible cultural or systemic re-modelling. The primary agents "neither express interests nor organize for their strategic pursuit" (p. 259). This is not to say that the aggregated effects of primary agents can have no impact at a cultural or system level, they can well generate aggregated and powerful impacts at a systemic level but the outcomes are a consequence of uncoordinated action and without stated aim.

The social interaction of programs like an EA implementation play out in an environment of corporate agents promoting the systemic state in question. Corporate agency thus shapes the context for primary agency. Yet primary agency also has aggregate effects as they unleash a range of environmental pressures and problems which may impact the aims that the corporate agent seeks. This is what Archer refers to as double morphogenesis – "…where agency undergoes transformation, acquiring new emergent powers in the very process of seeking to reproduce and transform structures" (p. 190). Corporate agency thus has two tasks with respect to the promotion of their goals: "the pursuit of its self-declared goals, as defined in a prior social context, and their continued pursuit in an environment modified by the responses of Primary Agency to the context which they confront" (p. 260).

The acceptance of EA is ultimately dependent on ensuring that primary agents accept its basic premise given the context within which they reside. As Horrocks (2009) suggests the recognition of a distinction between corporate and primary agents is useful in examining program implementation. For EA there are many cases where agents are ambivalent to EA and its underlying premise, yet their aggregate effects may well be significant in the ultimate rejection or acceptance. Those who have little understanding of IT and its strategic role will still need to be convinced of the ultimate benefits of restrictions to their current ways of doing. Presenting EA to such a group is a challenge. The program mechanisms must place a major focus on communication of the benefits in context, taking clear recognition of the cultural and ideational elements involved.

|  | Contradictions | | Complementarities | |
|---|---|---|---|---|
|  | **Necessary** | **Contingent** | **Necessary** | **Contingent** |
| **Situational logic** | correction | elimination | protection | opportunism |
| **CEP's: Cultural system** | syncretism | pluralism | systematization | specialization |
| **CEP's: Socio-cultural interaction** | unification | cleavage | reproduction | sectionalism |
| **SEP's: Structural system** | compromise | competition | integration | differentiation |
| **SEP's: Social interaction** | containment | polarisation | solidarity | diversification |

*Figure 3: Situational Logics at different strata (adapted from Archer, 1995 p. 303)*

# 4 Research Method

## 4.1 Objective, case study and data collection

The research objective is to understand the role of people in the EA implementation and its interplay with culture, social structure and the socio-cultural system by building a theoretical explanations of the role of people as the key element in EA implementation.

The case study has been conducted in university X – a large multi-campus institution serving local communities as well as a significant cohort of international students. The university has recognised that its present and future depends importantly on the institution's implementation of EA to deliver its mission and strategic priorities. Although the research in this case focused on a single case example the adoption of a focus on mechanisms suggests the arguments can be used more generally in other universities. As Stake (1994) (cited in Dobson (2001) argues this approach is in line with the argument that to complete a case study is not a methodological choice but a choice as to the object to be studied.

The research utilises a range of semi-structured interviews across various university stakeholders as the primary method of data collection. Interview questions and protocols are constructed to unearth perceived causal inferences, while direct observation also plays a role as the researcher is a passive participant in EA regular meetings, and reviews archival data.

The population targets interviewed were divided into two categories: (a) 'corporate agents' and 'primary agents' who affect and are affected by the EA implementation (the EA management roles ranging from the university IT Governance Committee, the CIO, the university business group to the lead enterprise architect), and (b) 'individual actors' who can be affected by the EA implementation (the EA end-users: lecturers, researchers, students; and staff of administrative centres, IT and the library). The EA end-users randomly selected for interview were from faculties, schools, students and service centres.

## 4.2 Morphogenetic approach as an analytical tool

MA as a central theme, attempts to identify the interplay between culture, structure and agency over time as the EA implementation ensues. The analysis also seeks to identify possible mechanisms: (a) the key program mechanisms that need to implemented to drive such large-scale architectural transformation, (b) the key social mechanisms that are in place in the organisation at the time of EA implementation.

In figure 4 below, different SEPs and CEPs exist at time 1 (T1) – these emergent organisational structural properties or powers of the structural system (S.S) and cultural system (C.S) define particular situational logics, which predispose agents towards specific course of action for the promotion of their interest. The form of situational logic represents the possible situational social mechanisms that shape the people opportunities and orientations. A stratified model of people (in figure 1) will help define how the EA program engages with the levels in different ways. Over Time 2 (T2) to time 3 (T3) examination of social/socio-cultural interaction (S-I/S-C) will ensue to understand how the different strata take up different courses of action within the EA program. Over the interaction stage people will develop their PEPs based on: (a) the available alternatives to the people involved, (b) the restrictions that govern the choice of alternatives, and (c) their evaluation of the possible consequences of the choices made (Archer, 2010). S-I/S-C interactions are seen as being structurally conditioned but never as structurally determined and agents themselves are seen to possess their own irreducible emergent powers (as seen in figure 2).

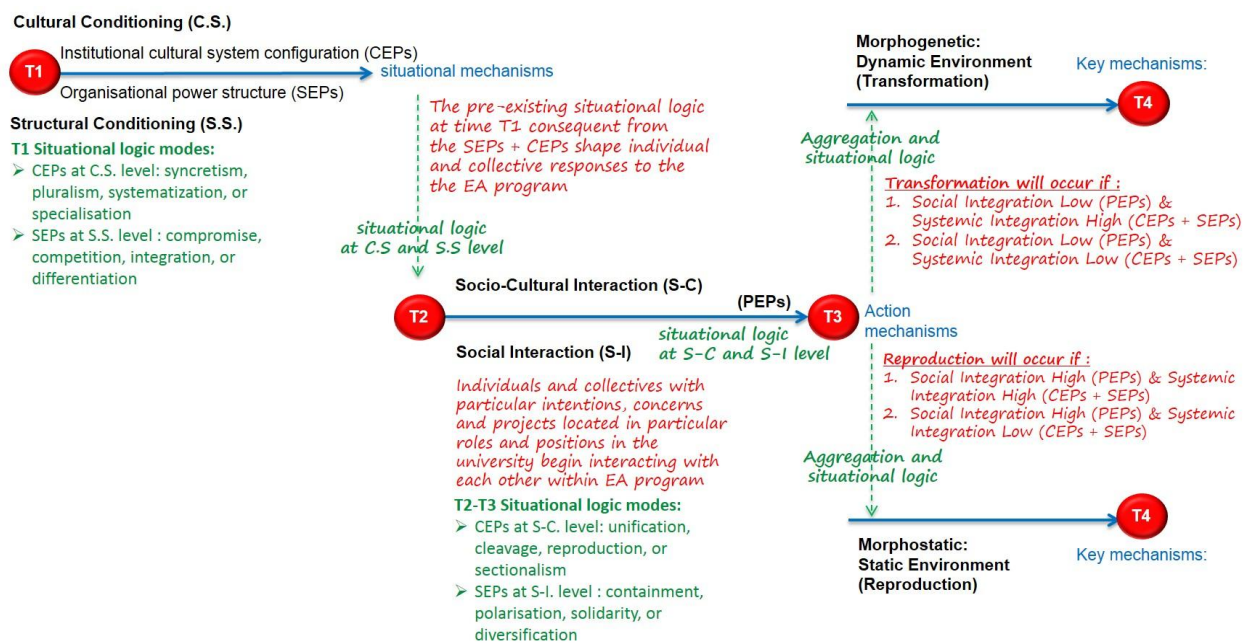

*Figure 4: MA as an explanatory/analytical framework (adapted from Archer, 1995, pp. 193, 218, 295, and 303)*

At time 4 (T4), as described by Luckett (2012, p. 324), analytic histories of particular contexts are required to explain the outcomes of social interaction which may involve structural and cultural change [morphogenesis: when the majority of university environment adopt the EA program – positive feedback predominate – to elaborate or change the social system's given form/structure] or reproduction [morphostatis: the majority of university environment reject the EA program – governed fundamentally by negative feedback – thus preserving the social system's given form/structure unchanged].

## 5   Conclusion

As the Chief Technology Officer (CTO) indicates the main purpose of EA is to communicate to stakeholders the benefits of EA; the CTO also sees 'people' issues as the primary factor in adoption:

> *(a) I think the main change from EA will be about governance and investment in technology - who makes the decisions? ... [The purpose of the EA program] is really educating people about the value of EA and getting them to understand it and then communicating and keeping the linkages between the EA and the business of the university;*

> *(b) I think the difference is in probably two things. One is the level of understanding about enterprise architecture and how choices about technology actually can have an impact on the outcome that you achieve from a business perspective. And it probably comes down to an individual level as cultural change things do. At that individual level, individuals' willingness to adapt and to actually see that there are lots of different ways of doing things... [The other is that] Some people will resist just because they want to resist and some of it is a lack of understanding and that seemingly loss of control-type thing. So I think they're the things that we need to address in involving people and communicating to people.*

Consistent with these arguments a member of the IT Governance Committee indicates the important need to communicate the value of EA:

> *(a) I think there is a real risk though, of it [EA] being presented at too high a level within the university and people just, again, not understanding it and then the perception is 'oh, that's something that IT does', and not understanding that they need to be involved in it, and if they're not involved in it there's going to be a disconnect between what potentially IT people or consultants think is the enterprise architecture and what the business actually needs. You're not going to have that joined. So it's incredibly important that there's a communication program in place and that the stakeholders here accept enterprise architecture and that they understand the importance of it.*

> *(b) ... [Some people] are possibly the ones that have the most to lose out of enterprise architecture. Maybe previously they had a bit more free rein to spend their IT budget on the things they thought were the most important, which wouldn't necessarily be the things that will deliver the most value to the university or didn't align with the strategic direction of the university, whereas now they might view it as an enterprise architecture roadmap for their business area or it might constrain what they would want to do because it would give the university reason to say 'no, you can't invest in that because it's not on your roadmap'.*

Given the key role that culture, structure and agency will play in EA implementation the benefit of a model incorporating them is self-evident. This paper describes some of the fertile possibilities of the morphogenetic approach for understanding how culture, structure and agency are potentially linked. It can provide help in understanding how an organisation's stakeholders might affect and be affected by EA implementation, and how they may react to the various EA impositions and possibilities. Application of Archer's representation of situational logic also can provide a useful high-level means for understanding the possible pathways that EA adoption can follow in other university environments.

## 6   References

Anderson, P. and Backhouse, G. 2009. "Unleashing EA: Institutional architectures and the value of joined up thinking," Bristol, United Kingdom: JISC Enterprise Architectures Group Pilot – Intelligent Content Ltd. http://www.jisc.ac.uk/enterprisearchitectures Retrieved: 04 February 2014.

Archer, M. 1975. *Problems of current sociological research: The work of the research committees of the International Sociological Association* (Editorial). Current Sociology, 23 (0011-3921)

Archer, M. 1979. *Social Origins of Educational Systems*, United Kingdom: London: Sage

Archer, M. 1989. 'Theory, culture and post-industrial society', *Sociologia: Revista di Scienza Sociali*, 1 n.s

Archer, M. 1995. *Realist Social Theory: The Morphogenetic Approach*. Cambridge, United Kingdom: Cambridge University Press, pp. 90–216.

Archer, M. 2010. "Routine, Reflexivity, and Realism," Washington, USA: 2010 American Sociological Association. *Sociological Theory* 28 (3), pp. 273–303.

Archer, M. 2013. *Social Morphogenesis*. New York, USA: Springer

Bente, S., Bombosch, U. and Langade, S. 2012. *Collaborative Enterprise Architecture: Enriching EA with Lean, Agile, and Enterprise 2.0 Practices*. Elsevier Inc, USA, pp. 35–36.

Bhaskar, R. 1978. *A Realist Theory of Science*. Harvester Press, Sussex, p. 25.

Dobson, P. J. (2001). Longitudinal case research: A critical realist perspective. Systemic practice and action research, 14 (3), pp. 283-296.

Gartner, Inc. 2009. Ten enterprise architecture pitfalls. http://www.gartner.com/newsroom/id/1159617 Retrieved: 22 June 2013.

Gengatharen, D., Standing, C., and Knight, S. 2009. "Knowledge management in an organisational climate of uncertainty and change: A longitudinal case study of an Australian university," *20th Australian Conference of Information Systems*, Melbourne, Australia.

Janssen, M. 2012. "Sociopolitical aspects of interoperability and enterprise architecture in e-government," *Social Science Computer Review 2012*, 30 (24), pp. 24-36. doi: 10.1177/0894439310392187

Luckett, K. 2012. Working with 'necessary contradictions': a social realist meta-analysis of an academic development programme review. http://dx.doi.org/10.1080/07294360.2011.631518 Retrieved: 04 June 2014, pp. 340 – 352.

Mason, K., Easton, G. and Lenney, P. 2013. Causal social mechanism: from the what to the why. *Journal of Industrial Marketing Management*. 42(3), pp. 347–355.

Raadt, B., Schouten, S. and Vliet, H. 2008. Stakeholder perception of enterprise architecture. *ECSA 2008, LNCS 5292*, (pp. 19–34). Berlin, Germany: Springer-Verlag Berlin Heidelberg.

## Copyright